\documentclass[twocolumn,showpacs,pra]{revtex4}
\usepackage{amssymb}
\usepackage{makeidx}
\usepackage{amsmath}
\usepackage{graphicx}
\usepackage{dcolumn}
\usepackage{bm}

\begin{document}

\title{Two-level system with broken inversion symmetry coupled to a quantum
harmonic oscillator}
\author{H.K. Avetissian}
\author{G.F. Mkrtchian}
\affiliation{Centre of Strong Fields Physics, Yerevan State University, 1 A. Manukian,
Yerevan 0025, Armenia}

\begin{abstract}
We study the generalized Jaynes-Cummings model of quantum optics at the
inversion-symmetry-breaking and in the ultrastrong coupling regime. With the help of
a generalized multiphoton rotating-wave approximation, we study the
stationary solutions of the Schr\"{o}dinger equation. It is shown that the
problem is reduced to resonant interaction of two position-displaced
harmonic oscillators. Explicit expressions for the eigenstates and
eigenvalues of generalized Jaynes-Cummings Hamiltonian are presented. We
exemplify our physical model with analytical and numerical considerations
regarding collapse and revivals of the initial population of a two-level
system and photon distribution function at the direct multiphoton resonant
coupling.
\end{abstract}

\maketitle



\section{Introduction}

Two-level system coupled to a quantum harmonic oscillator (e.g., a single
radiation mode) as a simple and tractable model has played a central role in
many branches of contemporary physics ranging from quantum optics to
condensed matter physics. In quantum optics it describes a two level atom
resonantly coupled to a single mode electromagnetic radiation \cite{Scully},
so called Jaynes-Cummings (JC) model \cite{Jaynes}. It accurately describes
trapped ion experiments for quantum informatics \cite{ion}. In condensed
matter physics we may include here Holstein model~\cite{Holstein}, graphene
in the magnetic field \cite{RB} or in the quantized single mode radiation
field \cite{Kib}, quantum dots coupled to photonic cavities \cite{Dots}, and
circuit quantum electrodynamics (QED) setups where superconducting qubits
are coupled to microwave cavities \cite{CirQED}. Even though the underlying
setups of mentioned systems are different, the physics is similar to Cavity
QED, where first experiments have been done toward the realization of JC
model \cite{RW}. Cavity QED can be divided into three coupling regimes:
weak, strong, and ultrastrong. For weak coupling atom-photon interaction
rate is smaller than the atomic and cavity field decay rates. In this case
one can manipulate by the spontaneous emission rate compared with its vacuum
level by tuning discrete cavity modes \cite{Weak}. In strong coupling
regime, when the emitter--photon interaction becomes larger than the
combined decay rate, instead of the irreversible spontaneous emission
process coherent periodic energy exchange between the emitter and the photon
field in the form of Rabi oscillations takes place \cite{CQED}. Thanks to
recent achievements in solid-state semiconductor \cite{sss} or
superconductor systems \cite{scs} one can achieve ultrastrong coupling
regime, where the coupling strength is comparable to appreciable fractions
of the mode frequency. In this regime new nonlinear phenomena are visible 
\cite{NL}. Besides, in these setups one can enrich the conventional JC model
including new interaction terms inaccessible in conventional Cavity QED
setups. One of the new factor which can be incorporated into the JC model is
an inversion-symmetry-breaking (ISB). Thus, in the conventional JC model, as
well as in the Rabi model with classical radiation field one assumes that
the diagonal matrix elements of the dipole moment operator are zero, that is
the states possess a certain spatial parity, and the levels are not
degenerated. Nevertheless, in various systems of interest, there is
intrinsic or extrinsic reasons for ISB. The inversion symmetry of a system
can be broken either by a system Hamiltonian or the stationary states may
not have this symmetry. As has been shown in Refs. \cite{B,M1}, when the
quantum system has permanent dipole moments in the stationary states, or the
level is degenerated upon orbital momentum there are new multiphoton effects
in the quantum dynamics of the system subjected to a strong laser field.
Furthermore, these systems have an advantage, which allows to generate
radiation with Rabi frequency \cite{M2} and moderately high harmonics by
optical pulses \cite{M3}. For the semiconductor version of JC model one can
achieve ISB by the asymmetric quantum dots \cite{AQD}. In the circuit QED
setups it appears naturally as a consequence of internal asymmetry. For flux
qubit potential landscape is reduced to a double-well potential \cite{scs},
for Cooper pair box ISB takes place at setup far from charge degeneracy
point \cite{MC}. Thus, it is of interest to study the consequence of ISB on
the quantized version of Rabi model, where multiphoton effects are expected
in the ultrastrong coupling regime.

In the present work we study the effect of ISB on the quantum dynamics of a
two-level system interacting with a quantized harmonic oscillator.
Particularly, we consider the consequences of the ISB on the eigenstates and
eigenenergies of generalized JC Hamiltonian, and on the dynamics of Rabi
oscillations, collapse and revival. It is shown that ISB substantially
alters the dynamics of the system compared with conventional JC one. Similar
to quasiclassical case \cite{M1} it is possible direct multiphoton
transitions, and as a consequence, there are Rabi oscillations with periodic
exchange of several photons between the emitter and the radiation (bosonic)
field. We consider ultrastrong coupling regime. Accordingly, the quantum dynamics
of the considered system is investigated using a resonant approximation.

The paper is organized as follows. In Sec. II the model Hamiltonian is
presented and diagonalized in the scope of a resonant approximation. In Sec.
III we consider temporal quantum dynamics of considered system and present 
corresponding numerical simulations. Finally, conclusions are given in Sec.
IV.

\section{Basic model Hamiltonian and dressed states picture}

Assuming here two level system with ISB coupled to quantum harmonic
oscillator, the model Hamiltonian can be written as%
\begin{equation*}
\widehat{H}=\hbar \omega \left( \widehat{a}^{+}\widehat{a}+\frac{1}{2}%
\right) +\frac{\hbar \omega _{0}}{2}\widehat{\sigma }_{z}
\end{equation*}%
\begin{equation}
+\hbar \left( -\lambda _{g}\widehat{\sigma }_{\downarrow }+\lambda _{e}%
\widehat{\sigma }_{\uparrow }+\lambda _{eg}\widehat{\sigma }_{x}\right)
\left( \widehat{a}^{+}+\widehat{a}\right) .  \label{H_m}
\end{equation}%
The first two terms in Eq. (\ref{H_m}) correspond to the free harmonic
oscillator of frequency $\omega $ and two level system with the transition
frequency $\omega _{0}$, respectively. The final term gives the interaction
between the oscillator and two level system. Creation and annihilation
operators, $\widehat{a}^{+}$and $\widehat{a}$, satisfy the bosonic
commutation rules, $\widehat{\sigma }_{x}$, $\widehat{\sigma }_{z}$ are
Pauli operators, $\widehat{\sigma }_{\uparrow }=\left( \widehat{I}+\widehat{%
\sigma }_{z}\right) /2$ and $\widehat{\sigma }_{\downarrow }=\left( \widehat{%
I}-\widehat{\sigma }_{z}\right) /2$ are projection operators and are the
result of ISB. These terms distinguish the systems being considered from
conventional JC model. At $\lambda _{g}=\lambda _{e}=0$, one will obtain
usual Hamiltonian for JC model (including also counter-rotating terms) with
coupling $\hbar \lambda _{eg}$. In the case of atoms/molecules and quantum
dots $\lambda _{g}$ and $\lambda _{e}$ correspond to mean dipole moments in
states of indefinite parity, while $\lambda _{eg}$ corresponds to transition
dipole moment. In the case of circuit QED see Refs. \cite{scs,MC}. Without
loss of generality we have assumed that ground and excited states have mean
dipole moments of opposite signs, and $\lambda _{e},$ $\lambda _{g}$ $\geq 0$%
.  
\begin{figure}[tbp]
\includegraphics[width=.48\textwidth]{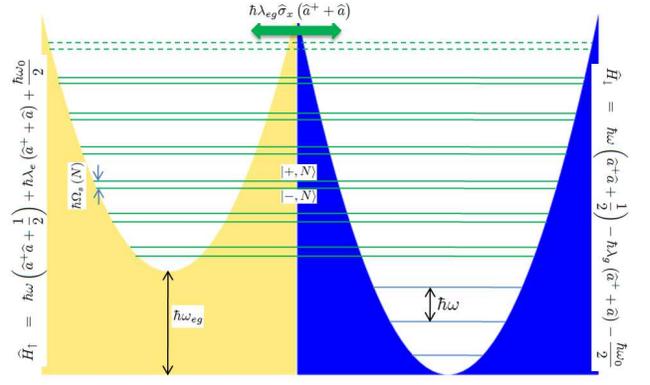}
\caption{Schematic illustration of the two coupled position-displaced
harmonic oscillators. In each well the eigenstates are displaced Fock
states. At the resonance, the energy levels starting from the ground state
of upper harmonic oscillators are degenerated. The coupling removes this
degeneracy, leading to symmetric and asymmetric entangled states. The
splitting of levels is defined by the vacuum multiphoton Rabi frequency.}
\end{figure}

At first we will diagonalize the Hamiltonian (\ref{H_m}), which is
straightforward in the dressed states picture. As JC model our model does
not admit exact analytical solution. One of the most powerful approximations
for the solution of JC model is the resonant or so called rotating-wave
approximation (RWA), which is valid at near-resonance $\left\vert \omega
_{0}-\omega \right\vert <<$ $\omega $ and weak coupling between the two
systems $\left\vert \lambda _{eg}\right\vert <<\omega _{0}$~\cite{Scully}.
For our model generalized multiphoton RWA is needed. The first step is to
rewrite Hamiltonian (\ref{H_m}) in the form 
\begin{equation}
\widehat{H}=\widehat{H}_{0}+\widehat{V},  \label{dec}
\end{equation}%
where $\widehat{H}_{0}=$ $\widehat{H}_{\uparrow }\otimes \widehat{\sigma }%
_{\uparrow }+\widehat{H}_{\downarrow }\otimes \widehat{\sigma }_{\downarrow }
$ represents two non-coupled position-displaced oscillators:%
\begin{equation}
\widehat{H}_{\uparrow }=\hbar \omega \left( \widehat{a}^{+}\widehat{a}+\frac{%
1}{2}\right) +\frac{\hbar \omega _{0}}{2}+\hbar \lambda _{e}\left( \widehat{a%
}^{+}+\widehat{a}\right) ,  \label{Ha}
\end{equation}%
\begin{equation}
\widehat{H}_{\downarrow }=\hbar \omega \left( \widehat{a}^{+}\widehat{a}+%
\frac{1}{2}\right) -\frac{\hbar \omega _{0}}{2}-\hbar \lambda _{g}\left( 
\widehat{a}^{+}+\widehat{a}\right) ,  \label{Hb}
\end{equation}%
and%
\begin{equation}
\widehat{V}=\hbar \lambda _{eg}\widehat{\sigma }_{x}\left( \widehat{a}^{+}+%
\widehat{a}\right)   \label{h2}
\end{equation}%
is the interaction part. Hamiltonians (\ref{Ha}) and (\ref{Hb}) admit exact
diagonalization. Schematic illustration of the two position-displaced
harmonic oscillators with coupling $\widehat{V}$ is given in Fig. 1. It is
easy to see that in each well the eigenstates are%
\begin{eqnarray}
| &\uparrow &,N^{(\lambda _{e})}\rangle \equiv |\uparrow \rangle \otimes
e^{-(\lambda _{e}/\omega )(\hat{a}^{\dag }-\hat{a})}|N\rangle ,  \notag \\
| &\downarrow &,N^{(\lambda _{g})}\rangle \equiv |\downarrow \rangle \otimes
e^{(\lambda _{g}/\omega )(\hat{a}^{\dag }-\hat{a})}|N\rangle ,  \label{eg1}
\end{eqnarray}%
with energies%
\begin{eqnarray}
E_{eN} &=&\frac{\hbar \omega _{0}}{2}+\hbar \omega (N+\frac{1}{2})-\hbar 
\frac{\lambda _{e}^{2}}{\omega },  \notag \\
E_{gN} &=&-\frac{\hbar \omega _{0}}{2}+\hbar \omega (N+\frac{1}{2})-\hbar 
\frac{\lambda _{g}^{2}}{\omega }.  \label{egen1}
\end{eqnarray}%
Hear $D\left( \alpha \right) =e^{\alpha (\hat{a}^{\dag }-\hat{a})}$ is the
displacement operator and quantum number $N=0,1,...$. The states $|\uparrow
\rangle $, $|\downarrow \rangle $ are eigenstates of $\hat{\sigma}_{z}$ and
the states $|N^{(\lambda _{e})}\rangle $, $|N^{(\lambda _{g})}\rangle $ are
position-displaced Fock states:%
\begin{eqnarray}
|N^{(\lambda _{e})}\rangle  &=&e^{-(\lambda _{e}/\omega )(\hat{a}^{\dag }-%
\hat{a})}|N\rangle =\sum_{M}I_{N,M}\left( \frac{\lambda _{e}^{2}}{\omega ^{2}%
}\right) |M\rangle ,  \notag \\
|N^{(\lambda _{g})}\rangle  &=&e^{(\lambda _{g}/\omega )(\hat{a}^{\dag }-%
\hat{a})}|N\rangle =\sum_{M}I_{M,N}\left( \frac{\lambda _{g}^{2}}{\omega ^{2}%
}\right) |M\rangle ,  \label{PDFock}
\end{eqnarray}%
where $I_{N,M}\left( \alpha \right) $ is the Lagger function and defined via
generalized Lagger polynomials $L_{n}^{l}\left( \alpha \right) $ as follows:%
\begin{eqnarray}
I_{s,s^{\prime }}\left( \alpha \right)  &=&\sqrt{\frac{s^{\prime }!}{s!}}e^{-%
\frac{\alpha }{2}}\alpha ^{\frac{s-s^{\prime }}{2}}L_{s^{\prime
}}^{s-s^{\prime }}\left( \alpha \right) =\left( -1\right) ^{s-s^{\prime
}}I_{s^{\prime },s}\left( \alpha \right) ,  \notag \\
L_{n}^{l}\left( \alpha \right)  &=&\frac{1}{n!}e^{\alpha }\alpha ^{-l}\frac{%
d^{n}}{d\alpha ^{n}}\left( e^{-\alpha }\alpha ^{n+l}\right) .  \label{Lag}
\end{eqnarray}%
Particularly, $|0^{(\lambda _{e})}\rangle $ and $|0^{(\lambda _{g})}\rangle $
are the Glauber or coherent states with mean number of photons $\lambda
_{e}/\omega $ and\ $\lambda _{g}/\omega $. Thus, we have two ladders shifted
by the energy: 
\begin{equation}
\hbar \omega _{eg}=\hbar \left( \omega _{0}+\lambda _{g}^{2}/\omega -\lambda
_{e}^{2}/\omega \right) .  \label{shift}
\end{equation}%
The coupling term (\ref{h2}) $\widehat{V}\sim \widehat{\sigma }_{x}$ induces
transitions between these two manifolds. At the resonance: 
\begin{equation}
\omega _{eg}-\omega n=\delta _{n};\ \left\vert \delta _{n}\right\vert
<<\omega   \label{res}
\end{equation}%
with $n=1,2,...$ the equidistant ladders are crossed: $E_{eN}\simeq E_{gN+n}$%
, and the energy levels starting from the ground state of upper harmonic
oscillators are nearly degenerated. The coupling (\ref{h2}) removes this
degeneracy, leading to symmetric and asymmetric entangled states. The
splitting of levels is defined by the vacuum multiphoton Rabi frequency. In
this case we should apply secular perturbation theory resulting:%
\begin{equation*}
|\alpha ,N\rangle =\left( C_{\downarrow }^{(\alpha )}|\downarrow
,N^{(\lambda _{g})}\rangle +C_{\uparrow }^{(\alpha )}|\uparrow ,\left(
N-n\right) ^{(\lambda _{e})}\rangle \right) ,
\end{equation*}%
\begin{equation}
E_{\alpha ,N}=\frac{1}{2}\left( E_{gN}+E_{eN-n}\right) +\alpha \sqrt{\frac{1%
}{4}\delta _{n}^{2}+\left\vert V_{N}\left( n\right) \right\vert ^{2}},
\label{dres}
\end{equation}%
where $\alpha =\pm $; $C_{\downarrow }^{(\alpha )}$ and $C_{\uparrow
}^{(\alpha )}$ are constant with ratio 
\begin{equation*}
C_{\uparrow }^{(\alpha )}/C_{\downarrow }^{(\alpha )}=V_{N}\left( n\right)
/\left( E_{\alpha ,N}-E_{eN-n}\right) ,
\end{equation*}
and transition matrix element is:%
\begin{equation*}
V_{N}\left( n\right) =\langle \downarrow ,N^{(\lambda _{g})}|\widehat{V}%
|,\uparrow ,\left( N-n\right) ^{(\lambda _{e})}\rangle 
\end{equation*}%
\begin{equation}
=\hbar \lambda _{eg}\left[ \frac{\lambda _{g}-\lambda _{e}}{\omega }-\frac{%
n\omega }{\lambda _{e}+\lambda _{g}}\right] I_{N-n,N}\left( \frac{\left(
\lambda _{g}+\lambda _{e}\right) ^{2}}{\omega ^{2}}\right) .  \label{Vn}
\end{equation}%
For the exact resonance, starting from the $N=n$ we have symmetric and
asymmetric entangled states%
\begin{equation}
|\pm ,N\rangle =\left( |\downarrow ,N^{(\lambda _{g})}\rangle \pm |\uparrow
,\left( N-n\right) ^{(\lambda _{e})}\rangle \right) /\sqrt{2}  \label{er}
\end{equation}%
with energies $E_{\pm ,N}=E_{gN}\pm \left\vert V_{N}\left( n\right)
\right\vert $, while for $N=0,1...n-1$, we have eigenstates $|\downarrow
,N^{(\lambda _{g})}\rangle $ and energy $E_{gN}$. For the conventional JC
model there is a selection rule: $V_{N}\left( n\right) \neq 0$ only for $%
n=\pm 1$. This also follows from Eq. (\ref{Vn}) in the limit $\lambda
_{e},\lambda _{g}\rightarrow 0$. That is why in that case only one photon
Rabi oscillations takes place. In our model due to ISB there are transition
with arbitrary $n$ giving rise to multiphoton coherent transitions. Besides,
at the $\lambda _{g}\neq 0$ in the ground state $|\downarrow \rangle \otimes 
$ $|0^{(\lambda _{g})}\rangle $ bosonic field is in the coherent state. The
solutions (\ref{dres}) are valid at near multiphoton resonance $\omega
_{eg}\simeq n\omega $ and weak coupling $\left\vert V_{N}\left( n\right)
\right\vert <<\omega .$ The latter condition implies that for the
multiphoton resonant transitions, systems with large dipole moments ($%
\left\vert \lambda _{e}+\lambda _{g}\right\vert >>\left\vert \lambda
_{eg}\right\vert $) are preferable.

\section{Multiphoton Rabi Oscillations}

Let us now consider the quantum dynamics of the two-level system and
harmonic oscillator starting from an initial state, which is not an
eigenstate of the Hamiltonian (\ref{H_m}). This is of particular interest
for applications in quantum information processing. Assuming arbitrary
initial state $|\Psi _{0}\rangle $ of a system, then the state vector for
times $t>0$ is just given by the expansion over dressed state basis obtained
above:%
\begin{equation*}
|\Psi \left( t\right) \rangle =\sum_{N=0}^{n-1}\langle \downarrow
,N^{(\lambda _{g})}||\Psi _{0}\rangle e^{-\frac{i}{\hbar }%
E_{gN}t}|\downarrow ,N^{(\lambda _{g})}\rangle 
\end{equation*}%
\begin{equation}
+\sum_{\alpha =\pm }\sum_{N=n}^{\infty }\langle \alpha ,N||\Psi _{0}\rangle
e^{-\frac{i}{\hbar }E_{\alpha ,N}t}|\alpha ,N\rangle .  \label{DWF}
\end{equation}%
For concreteness we will consider two common initial conditions for harmonic
oscillator: the Fock state and the coherent state. We will calculate the
time dependence of the two level system population inversion $W_{n}\left(
t\right) =$ $\langle \Psi \left( t\right) |\widehat{\sigma }_{z}|\Psi \left(
t\right) \rangle $ at the exact $n$-photon resonance (\ref{res}) $\delta
_{n}=0$. For the field in the Fock sate and two level system in the excited
state $|\Psi _{0}\rangle =|\uparrow ,0\rangle $, we have%
\begin{equation}
W_{n}\left( t\right) =\sum_{N=0}^{\infty }I_{N,0}^{2}\left( \frac{\lambda
_{e}^{2}}{\omega ^{2}}\right) \cos \left( \Omega _{N+n}\left( n\right)
t\right) ,  \label{W1}
\end{equation}%
where $\Omega _{N}\left( n\right) =2\left\vert V_{N}\left( n\right)
\right\vert /\hbar $ is the multiphoton vacuum Rabi frequency. For $\lambda
_{e}^{2}<<\omega ^{2}$ the main contribution in the sum (\ref{W1}) comes
from the first term: $W_{n}\left( t\right) \simeq \cos \left( \Omega
_{n}\left( n\right) t\right) $, which corresponds to Rabi oscillations with
periodic exchange of $n$ photons between the two-level system and the
radiation (bosonic) field. 

Finally we turn to the case in which a two-level system begins in the ground
state, while oscillator prepared in a coherent state with a mean excitation
(photon) number $\overline{N}$. From Eq. (\ref{eg1}) follows that such state
can be represented as $|\Psi _{0}\rangle =|\downarrow \rangle \otimes
|0^{(\lambda _{e}^{\prime })}\rangle $, where $\lambda _{e}^{\prime }=%
\overline{N}\omega $. Taking into account Eqs. (\ref{PDFock}) and (\ref{Lag}%
), for population inversion we obtain%
\begin{equation}
W_{n}\left( t\right) =-1+2\sum_{N=n}^{\infty }I_{N,0}^{2}\left( \rho \right)
\sin ^{2}\frac{\Omega _{N}\left( n\right) t}{2},  \label{W2}
\end{equation}%
where $\rho =\left( \overline{N}+\lambda _{g}/\omega \right) ^{2}$. In this
case we have collapse and revival phenomenon of the multiphoton Rabi
oscillations.

In this section, we also present numerical solutions of the time dependent
Schr\"{o}dinger equation with the full Hamiltonian (\ref{H_m}) in the Fock
basis considering up to $N_{\max }=200$ excitations. The set of equations
for the probability amplitudes has been solved using a standard fourth-order
Runge--Kutta algorithm \cite{27}.

\begin{figure}[tbp]
\includegraphics[width=.47\textwidth]{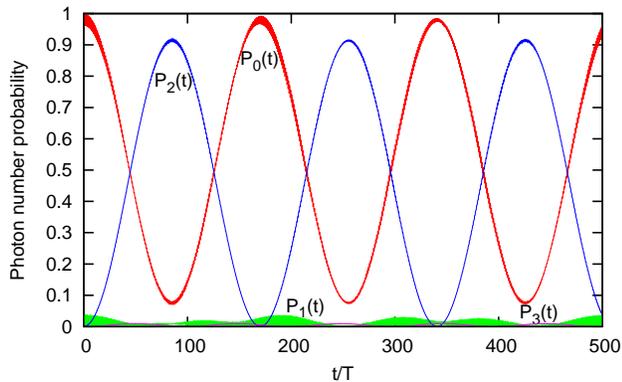}
\caption{(Color online) Photon number probability $P_{N}\left( t\right) $
(in arbitrary units) as a function of time (in units of oscillator period $%
T=2\protect\pi /\protect\omega $) at the two photon resonance. The coupling
parameters are $\protect\lambda _{eg}/\protect\omega =0.02$,$\ \protect%
\lambda _{g}/\protect\omega =0$, and $\protect\lambda _{e}/\protect\omega %
=0.1$. }
\end{figure}

\begin{figure}[tbp]
\includegraphics[width=.47\textwidth]{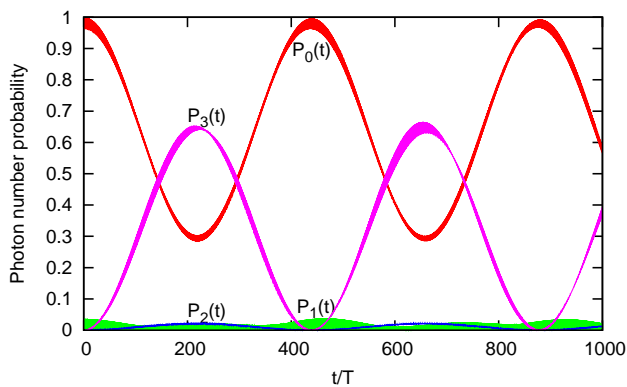}
\caption{(Color online) Same as Fig. 2 but for three-photon resonance and
coupling parameters - $\protect\lambda _{eg}/\protect\omega =0.02$,$\ 
\protect\lambda _{g}/\protect\omega =-0.1$, and $\protect\lambda _{e}/%
\protect\omega =0.1$.}
\end{figure}

\begin{figure}[tbp]
\includegraphics[width=.48\textwidth]{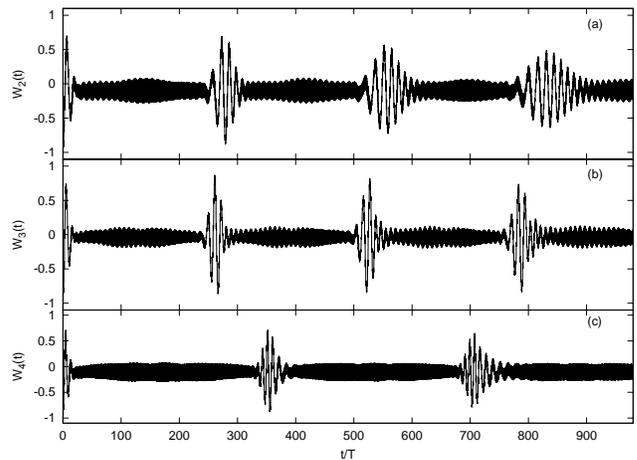}
\caption{ Collapse and revival of the multiphoton Rabi oscillations. Two
level system population inversion is shown with the field initially in a
coherent state. (a) Two-photon resonance with coupling parameters $\protect%
\lambda _{eg}/\protect\omega =0.02$,$\ \protect\lambda _{g}/\protect\omega %
=0 $, $\protect\lambda _{e}/\protect\omega =0.1$ and mean photon number $%
\overline{N}=20$. (b) Three-photon resonance with parameters - $\protect%
\lambda _{eg}/\protect\omega =0.02$,$\ \protect\lambda _{g}/\protect\omega %
=-0.1$, $\protect\lambda _{e}/\protect\omega =0.1$ and mean photon number $%
\overline{N}=30$. (c) Same as (b) but for four-photon resonance and $%
\overline{N}=60$. }
\end{figure}

\begin{figure}[tbp]
\includegraphics[width=.46\textwidth]{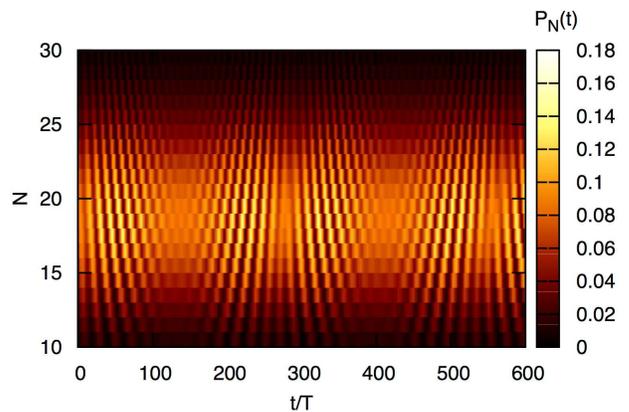}
\caption{(Color online) Density plot of photon number probability
distribution $P_{N}\left( t\right) $ (in arbitrary units) as a function of
photon number and time (in units of oscillator period $T=2\protect\pi /%
\protect\omega $) corresponding to setup of Fig. 4(a).}
\end{figure}
In Figs. (2) and (3) photon number probability%
\begin{equation*}
P_{N}\left( t\right) =\langle \uparrow ,N||\Psi \left( t\right) \rangle
\langle \Psi \left( t\right) |\uparrow ,N\rangle 
\end{equation*}%
\begin{equation}
+\langle \downarrow ,N||\Psi \left( t\right) \rangle \langle \Psi \left(
t\right) |\downarrow ,N\rangle   \label{Pn}
\end{equation}
as a function of time is shown for two and three photon resonances. For an
initial state we assume two level system in the excited state and the field
in vacuum state - $|\uparrow \rangle \otimes |0\rangle $. As is seen only
resonant multiphoton Fock states are excited. In Fig. 4 we show collapse and
revival of the multiphoton Rabi oscillations. Two level system inversion $%
W_{n}\left( t\right) $ is shown with the field initially in a coherent state
at two, three, and four photon resonances for different mean photon numbers.
The consequence of collapse and revival of the multiphoton Rabi oscillations
on the photon number probability distribution (\ref{Pn}) is shown in Fig.
(5). We see that numerical simulations are in agreement with analytical
treatment in the multiphoton RWA and confirm the revealed physical picture
described above.

\section{Conclusion}

We have presented a theoretical treatment of the quantum dynamics of a
two-level system with ISB interacting with a quantized harmonic oscillator
in the ultrastrong coupling regime. With the help of a resonant approach, we have
solved the Schr\"{o}dinger equation
and obtained simple analytical expressions for the eigenstates and
eigenenergies. The obtained results show that the effect of ISB on the
quantum dynamics is considerable. For the $n$-photon resonance in addition
to $n$ non-entangled states we have symmetric and asymmetric entangled
states of a two level system and position-displaced Fock states. The ground
state is a not entangled, but the bosonic field may be in a coherent state.
We have also investigated the temporal quantum dynamics of considered system
and showed that similar to one-photon case due to ISB it is possible Rabi
oscillations, collapse and revival of initial population with periodic
multiphoton exchange between the two-level system and the radiation field.
The proposed model may have diverse applications in Cavity QED experiments,
especially in the variant of circuit QED, where the considered parameters
are already accessible.

\begin{acknowledgments}
This work was supported by State Committee of Science (SCS) of Republic of
Armenia (RA), Project No. 13RF-002.
\end{acknowledgments}

\end{document}